\documentstyle[aps,pre,epsf,multicol]{revtex}
\begin{document}
\draft
\title{Anomalous Roughness, Localization, and Globally Constrained 
Random Walks}
\author{Jae Dong Noh$^{1,}$ \cite{JD}, Hyunggyu Park$^{1}$, Doochul Kim$^{2}$,
and Marcel den Nijs$^{3}$}
\address{$^{1}$Department of Physics, Inha University, Inchon 402-751,
Korea \\
$^{2}$School of Physics, Seoul National University, Seoul 151-742,
Korea \\
$^{3}$Department of Physics, University of Washington, Seattle,
WA 98195-1560, U.S.A.}
\date{\today}
\maketitle

\begin{abstract}
The scaling properties of a random walker subject to the global constraint
that
it needs to visit each site an even number of times are determined.
Such walks are realized in the equilibrium state of one dimensional surfaces
that are subject to dissociative dimer-type surface dynamics.
Moreover, they can be mapped onto unconstrained random walks on a 
random surface,
and the latter corresponds to a non-Hermitian random free fermion model
which
describes electron localization near a band edge.
We  show analytically that the dynamic exponent of this random walk  is
$z=d+2$ in  spatial dimension $d$. This explains the anomalous
roughness, with exponent $\alpha=1/3$, in one dimensional
equilibrium surfaces with dissociative dimer-type dynamics.
\end{abstract}


\begin{multicols}{2}

\section{introduction}

Random walks provide the key to the scaling properties of many 
physical phenomena.
Some examples are: surface diffusion, wave packet spreading in 
quantum mechanics,
equilibrium commensurate-incommensurate phase transitions in 
physisorbed monolayers
on substrates, and one dimensional (1D) interfaces  in equilibrium
and in stationary growing states~\cite{Fisher84,Bouchaud,denNijs88}.
In its elementary formulation, a particle  moves
through $d$ dimensional space by  jumping  during each discrete
time step over a displacement vector $\vec x$,
according to a given (localized) probability distribution $W(\vec x)$.
The fluctuations in the position,
${\vec n}_t = \sum_{t'=1}^t {\vec x}_{t'}$,
after $t$ time steps, obey the scaling form
\begin{equation}\label{diffusion}
\Delta {\vec  n}_t \equiv \langle | {\vec n}_t -
\langle {\vec n}_t \rangle|^2 \rangle^{1/2} \sim t^{1/z}
\end{equation}
and the spatial probability distribution $P(\vec n,t)$ develops
at large time scales into the Gaussian form
\begin{equation}
P({\vec n},t) = \frac{1}{(4\pi Dt)^{d/z}}
\exp\left[ -{|{\vec n}|^2}/{4Dt^{2/z}}\right] ~,
\end{equation}
which is invariant under the scale transformation $P\to b^{d} P$,
$t\to b^z  t$ and $\vec n\to b\vec n$.
$z$ is the so-called dynamic exponent, and has the simple value
$z=z_{rw}=2$ in all dimensions, as is obvious from
the linear character of the underlying diffusion equation.
Scale invariance is generic to many other dynamic processes as well,
but with typically  non-trivial values for $z$.
Kardar-Parisi-Zhang (KPZ)  type surface growth~\cite{KPZ}, and
statistical population dynamics like
directed percolation and directed Ising type processes
are examples of this~\cite{Hinrichsen_Park_review}.
In such processes random walk (diffusion) arguments
still play a central role, e.g., $z_{rw}$ typically appears
within molecular field type approximations.

The scaling properties of stochastic processes can be classified
in so-called universality classes, according to the symmetries and
conservation laws of the underlying dynamic rules.
In analogy with equilibrium critical phenomena,
dynamic critical exponents are universal quantities,
that are insensitive to varying details of the dynamic rule.
For example, all random walks have  $z=z_{rw}=2$, irrespective of the
shape of the displacement distribution $W(\vec x)$.
To break out of the $z_{rw}$ straight jacket, something
more drastic has to change.
Examples of anomalous diffusion are:
Levi flights, where the typical length of the displacement is divergent;
correlated random walks, where the steps have long range temporal
correlations~\cite{Bouchaud}; walks in a
quenched randomness environment, like
polymers in disordered media~\cite{Kardar_Zhang,JMKIM};
and also diffusion on a one dimensional lattice,
where the particles can move only in dimer form ~\cite{Dhar}.
The latter leads to an infinite
number of conserved quantities, and
density auto-correlation functions that decay as
power-laws with anomalous exponents.

In this paper, we investigate  the scaling properties of
so-called  even-visiting random walks~(EVRW) on a $d$ dimensional
hypercubic lattice.
During each time step $(0\leq t'<t)$ the random walker
hops to one of its nearest neighbor sites with equal probability.
However, unlike normal random walks, it is required to
visit every site an even number of times before the walk terminates 
at time $t$.
This gives rise to anomalous scaling.
The even-visiting condition  imposes a (mod 2 type) global constraint on
the motion of the random walker,
which it can satisfy only through correlated movements.
The origin and nature of this type of anomalous scaling
is completely different from those in the examples mentioned above.

This study of EVRW's is complementary to our recent work on
dissociative dimer-type surface growth in one 
dimension~(1D)~\cite{Noh_Park_MdN}.
The surface grows and erodes by the deposition and evaporation of dimers
only.
Those dimers dissociate while on the surface (but do not diffuse)
such that each monomer can arrive and depart  with a different partner.
This growth rule implies that the number of particles at each height level
is globally (but not locally) conserved mod $2$. Compared to conventional
(monomer deposition type) surface roughness,
it imposes a global constraint on
the fluctuations of the surface and leads to anomalous equilibrium
roughness.
The mod 2 conservation of the particle number in dimer growth is
equivalent to the even-visiting constraint in random walks, and 
the anomalous  surface roughness is linked to the anomalous  scaling 
of the EVRW.

There exists a second completely different type of application of EVRW's.
The time evolution operator of the random walk can be cast in the
form of quantum mechanical non-interacting electrons moving
in a random medium.
The global EVRW constraint translates into spatial and directional
randomness
of the hopping amplitudes and a non-Hermitian random 
Hamiltonian~(Sec.~\ref{sec:SEVRW}).
The spectral properties of such Hamiltonians are a topic of growing
research,
in particular their localization-delocalization transition
aspects~\cite{Hatano_Nelson,Feinberg_Zee}.
Our EVRW scaling involves only one electron and therefore relates to 
the scaling
of the electron eigenstates near the bottom of the (almost empty) energy
band.
Those edge states have so-called  Lifshitz tails, with  essential
singularities
in the disorder-averaged density of states~\cite{Luttinger}.
Our study of EVRW's elucidates the nature of the edge states.

The EVRW problem naturally generalizes to Q-visiting random walks~(QVRW)
with a constraint that the number of visits to each site must be
multiple of Q. Diffusive motion of the QVRW describes the
stationary state roughness of dissociative Q-mer growth, where
a surface grows~(erodes) by the deposition~(evaporation) of a  Q-mer.
This conserves the number of particles at each height level modulo Q,
which corresponds to the Q-visiting constraint.
In our discussion we will focus mostly on EVRW and dissociative dimer 
type growth,
but most results
are easily extended to QVRW  and the scaling properties turn out to be
independent of Q.

This paper is organized as follows.
In Sec.~\ref{sec:dimer}, we
review one dimensional dissociative dimer type 
dynamics~\cite{Noh_Park_MdN}, and
present additional numerical results showing the anomalous roughness of the
equilibrium surface.

In Sec.~\ref{sec:evrw}, we map this dimer growth model
onto the 1D EVRW and present numerical results for the 
anomalous diffusion
in 1D EVRW's. The finite size scaling type exact enumeration
and Monte Carlo simulation results suggest that the dynamic exponent
of the 1D EVRW is equal to $z\simeq 3.0$.
We also devise an analytic scaling argument, a healing time argument, that
predicts that the dynamic exponent is equal to  $z=d+2$ in $d$ dimensions.
The healing time argument presumes the existence of
a crossover time scale $\tau_{\text{free}}\sim t^{2/(d+2)}$,
within which the random walker is not influenced by the global even-visiting
constraint imposed at time $t$. We numerically confirm that
such a time scale is present in  $d=1$ and $d=2$.

In Sec.~\ref{sec:Ising}, we embed the  EVRW into an Ising spin environment.
A $S_{\vec n}=\pm 1$ spin is assigned to every site.
Each of them points up at time $t=0$.
Next, the particle moves as in a conventional unconstrained random walk,
but the Ising spin at site $\vec n$ flips with probability $e$ (or not
with probability $f=1-e$) each time the particle visits that site.
The global EVRW constraint can now be represented by the
requirement  that all spins must be pointing up again at time $t$, i.e., by
projecting out from all conventional RW configurations those that leave
spins pointing down.
The even-visiting walks condition obviously requires that $e=1$,
but it is advantageous to proceed with generalized
values $0\leq e\leq 1$. We call this generalized version the stochastic
even-visiting random walk (SEVRW).
Next, we construct the time evolution operator of the Ising spins
and the random walker.
The spin part is easily diagonalized and the result has the form
of a non-Hermitian single particle  quantum Hamiltonian with quenched 
randomness.
The latter can also be interpreted as the transfer matrix
(thermal equilibrium) of a polymer
fluctuating in an environment with randomly placed  defect-lines.

In Sec.~\ref{sec:walls} we focus on one special point in the SEVRW model,
$e=f=1/2$. There the 1D model is easily  soluble. 
The dynamic exponent is exactly equal to $z=3$.
This point plays the role of stable fixed point in the sense
of renormalization transformations in the EVRW phase diagram.
The EVRW, although at the border, belongs to its basin of attraction.

In Sec.~\ref{sec:Lifshitz} we discuss the same issues as in 
Sec.~\ref{sec:walls},
but now in the framework of the non-Hermitian quantum Hamiltonian.
The anomalous dynamic exponent $z=d+2$ relates to
so-called  Lifshitz tails in the density of states near the edge of
the energy band.
We conclude with a brief summary and discussion,
in Sec.~\ref{sec:summ_diss}.

\section{Dissociative Dimer Surface Dynamics}\label{sec:dimer}

\subsection{surface roughness}

Equilibrium and non-equilibrium properties of 1D interfaces like crystal
surfaces have attracted considerable interest~\cite{Barabasi}.
Interfaces display intriguing scaling properties and their growth
dynamics is well understood in terms of a few universality classes.
KPZ growth is one of the examples~\cite{KPZ}.
An interesting quantity associated with
interface roughness is the averaged width, $W$, i.e., the standard deviation
of the interface height $h_l(t)$ ($l=1,\ldots,L$):
\begin{equation}
W(L,t)^2 = \left\langle \frac{1}{L} \sum_{l=1}^L h_l(t)^2 -
\left(\frac{1}{L}\sum_{l=1}^L h_l(t)\right)^2 \right\rangle  \ ,
\end{equation}
where $L$ is the substrate size. The width satisfies the dynamic scaling
relation
\begin{equation}
W(L,t) = L^\alpha f(t / L^{z_W}) \ ,
\end{equation}
where the scaling function $f(x)$ behaves as
\begin{equation}
f(x) \sim \left\{
\begin{array}{c}
x^{\beta} \quad \mbox{for} \quad x\ll 1 \\
\mbox{const.} \quad \mbox{for} \quad x\gg 1 \ .
\end{array}
\right.
\end{equation}
The stationary state roughness exponent $\alpha$
and dynamic exponent $z_W=\alpha/\beta$ are universal quantities.

In most growth models the structural properties of the
depositing~(evaporating) particles are explicitly
or implicitly presumed to be geometrically featureless monomers.
Nevertheless, the geometric features of the particle shapes
can strongly affect the  growth dynamics  and the stationary state
scaling properties~\cite{Noh_Park_MdN}.

\subsection{dimer dynamics}

Consider a crystal built from atoms of type $X$. Assume  that
deposition  and  evaporation takes  place in contact with a gas where
the atoms only appear in molecular dimer form $X_2$, and assume that 
such dimers
can only land and depart from the interface when aligned with the surface.
Upon deposition  a dimer attaches to two nearest neighbor surface
sites and  loses its  dimer character.
Upon evaporation, two  nearest neighbor surface atoms form a dimer and
depart from the crystal. This dissociative character of the dimers
is the essential feature leading to the anomalous surface roughness
in the equilibrium state.

\begin{figure}
\centerline{\epsfxsize=84 mm \epsfbox{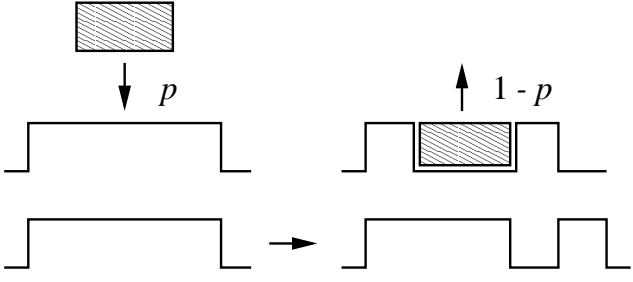}}
\vskip 10 true pt
\caption{The upper panels show the deposition and evaporation of a dimer.
The lower panels show diffusion of a monomer.}
\label{growth_rule}
\end{figure}

We describe  the 1D  surface configurations  in terms  of integer
height  variables subject to  the  so-called restricted
solid-on-solid~(RSOS) constraint,  $h_l-h_{l+1}=0,\pm 1$.
The dynamic rule is as follows.
First, select at  random a bond  $(l,l+1)$. If the  two
sites are not at the  same height, no  evaporation nor deposition
takes place.  If the two sites  are at the same  height, deposition
of a dimer covering both sites  is attempted with probability $p$,
or  evaporation of  a dimer with probability  $q=1-p$~(see
Fig.~\ref{growth_rule}). Processes
are  rejected if  they would  result  in a  violation of  the  RSOS
constraint.

Surfaces growing according to such
dissociative dimer dynamic rules behave fundamentally different from 
those following
monomer-type growth rules.
The latter, irrespective   of  being  in   equilibrium  or in
a stationary growing  state,  display, with only
a  few  very notable exceptions,  the universal roughness exponent
$\alpha=1/2$; as exemplified in the
Edwards-Wilkinson (EW)~\cite{EW}  and  the KPZ~\cite{KPZ} 
universality classes.
The  universal value of $\alpha$  is understood from a random walk
argument. To be precise, a 1D surface can be mapped on the time trajectory
of a particle in 1D by identifying the height $h_l$
at each site $l$ with the particle position $n_t$ at time  $t=l$.
The steps in 1D surfaces  are uncorrelated beyond a definite 
correlation length.
Therefore the particle performs a random walk with displacement fluctuations
$|n_t-n_{t'}| \sim (t-t')^{1/z_{rw}}$ at large time scales.
This  yields the value of the stationary state roughness exponent
$\alpha=1/z_{rw}=1/2$.

Dissociating dimer growth circumvents the  random
walk argument by means of a novel type of non-local topological
constraint. The dimer aspect requires  that the number of particles
at every surface height level must be conserved modulo 2.
The dissociative character of the dimers
transforms this into a non-local global feature.
This leads to various interesting phenomena.
In equilibrium, the surface is rough but with anomalous scaling
exponents~\cite{Noh_Park_MdN,Grynberg}.
Out of equilibrium, while growing or evaporating, it always 
facets~\cite{Noh_Park_MdN}.
Moreover, when the model is extended by introducing a so-called reduced
digging
probability at flat segments, towards a directed Ising type 
roughening transition
in the extreme no-digging limit, the roughness becomes even more 
complex~\cite{Noh_Park_MdN,Hinrichsen}.
The non-equilibrium faceting aspects are already well
documented in Ref.~\cite{Noh_Park_MdN}. Here we focus on
the anomalous equilibrium roughness.

\subsection{anomalous equilibrium roughness}

At $p=q$ the above dynamic rule satisfies the detailed balance 
condition and the
stationary state distribution is a genuine Gibbs type equilibrium state.
We study the dynamic scaling of the surface width via Monte Carlo 
(MC) simulations.
The crystal size $L$ is even, with periodic boundary conditions,
$h_{L+l}=h_l$,
and we use as initial condition a flat surface, $h_l = 0$ for all $l$.
The surface width is measured and averaged
over $N_s$ independent MC runs, ranging from $N_s= 5000$ for $L=2^{5}$ to
$N_s= 500$ for $L=2^{10}$.

\begin{figure}
\centerline{\epsfxsize=84 mm  \epsfbox{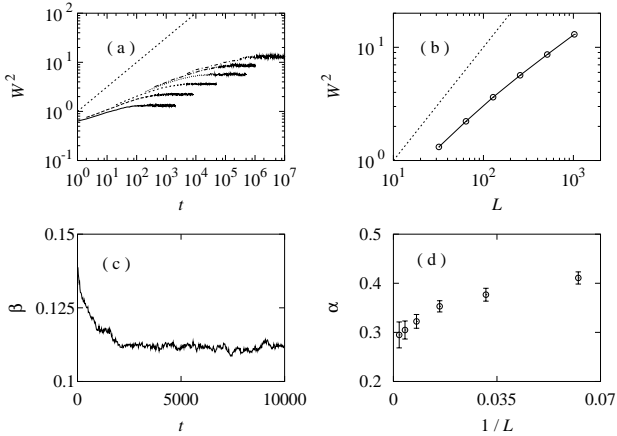}}
\vskip 10 true pt
\caption{Monte Carlo results for the dimer model.
(a) Time dependence of the surface width for
$L=32,
\ldots,1024$ from bottom to top. The straight line has slope
$2\beta_{EW}=1/2$.
(b) Saturated surface width. The straight line has slope $2\alpha_{EW}=1$.
(c) and (d) Effective values for $\beta$ and $\alpha$. }
\label{fig:width}
\end{figure}

The results are shown in Figs.~\ref{fig:width} (a) and (b).
The surface width does not obey monomer growth type EW scaling
with $\alpha=1/2$ and $\beta=1/4$.
The dimer surface width saturates slower $(\beta<1/4)$ and is
definitely less rough in equilibrium ($\alpha<1/2$). Notice the
large corrections to finite size scaling of the width in both
the temporal and spatial domains.
These prevent us from obtaining
accurate values for the  exponents $\alpha$ and $\beta$ from simple 
log-log type plots
of the width versus  $t$ and $L$.
Instead, we define effective exponents
\begin{equation}\label{alpha_eff}
\alpha(L) \equiv \ln [ W(mL,\infty)/W(L,\infty) ] / \ln m
\end{equation}
and
\begin{equation}\label{beta_eff}
\beta(t) \equiv \ln [ W(L,m t) / W(L,t) ] / \ln m \ ,
\end{equation}
where $m$ is arbitrary (we choose $m=2$) and
$W(L,\infty)$ denotes the saturated width.
For $\alpha(L)$, we use  data for $L=2^5,\ldots,2^{10}$, and for
$\beta(t)$, the data at $L=2^{13}$ at times shorter than $t<10^5$ where
finite size effects are still invisible.
The results are shown in Figs.~\ref{fig:width} (c) and (d).
We estimate
\begin{equation}\label{alpha_beta}
\alpha =  0.29(4) , \quad \beta = 0.111 (2) \ .
\end{equation}
and  $z_W \simeq 2.6(5)$, since $z_W=\alpha/\beta$. The
exponents are definitely different from those of ordinary equilibrium rough
interfaces but the precise values remain uncertain.

The mod 2 non-local conservation of particle number is clearly
the most promising candidate for being the origin of the anomalous 
scaling behavior;
as  confirmed  in the following sections.
However there exist additional more local conserved quantities in the 
dimer dynamics.
When a dimer desorbs or adsorbs, the surface heights at two nearest neighbor
sites change by one unit simultaneously. This implies conservation
of the anti-Bragg, $k = \pi$, Fourier component of the surface height
\begin{equation}
\tilde{h}_k \equiv \frac{1}{\sqrt{L}} \sum_{l=1}^L e^{-ikl} h_l \ .
\end{equation}
In other words, the dynamics is not ergodic;
surface configurations with different values of $\tilde{h}_{k=\pi}$ are
dynamically disconnected.
Therefore the scaling properties may depend on the initial condition.
Such types of effects are studied in Ref.~\cite{Grynberg} in the context of
dissociative $k$-mer growth in body-centered solid-on-solid
type models, $h_l-h_{l+1} = \pm 1$.

\subsection{surface diffusion}

In our model the particles do not diffuse along the surface.
In actual experimental settings, surface diffusion cannot be ignored.
The $k=\pi$ broken ergodicity is restored by diffusion, but
the mod 2 conservation is preserved as long as
diffusion across steps is forbidden.
Such jumps to higher and lower levels are suppressed
by  so-called Schwoebel barriers~\cite{Politi_etal}.
This means that the anomalous surface roughness discussed here
can be observed at time scales smaller than the characteristic time
associated with jumps across steps, provided the other time scales
are short (high surface deposition rates).

To test the robustness of anomalous dimer roughness
and to verify the essential role of the global mod 2
particle conservation at each height level, we
add to the dimer growth model diffusion of surface atoms within terraces.
The surface is again described by integer height variables
$h_l$, subject to the RSOS constraint and
periodic boundary conditions. The dynamic rule is as follows.
Select at random a bond $(l,l+1)$, and attempt with equal
probability:  a dimer deposition or evaporation just like above;
or a monomer jump from site $l$ to one of its nearest neighbor
sites.
The move is rejected if it would result in a violation of the RSOS
constraint. Since the RSOS condition is
imposed at every stage, jumps across steps are automatically forbidden.

\begin{figure}
\centerline{\epsfxsize=84 mm  \epsfbox{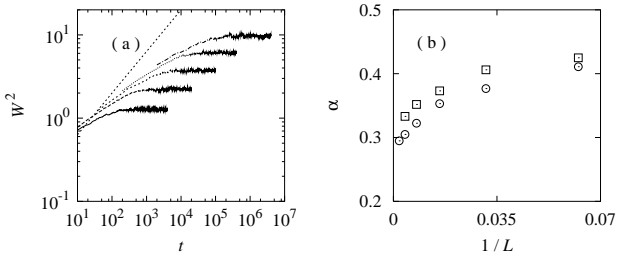}}
\vskip 10 true pt
\caption{(a) Time dependence of the surface width of the dimer
model with diffusion, for $L=32,\ldots, 512$ from bottom to top. The
straight
line has slope  $2\beta_{EW}=1/2$. (b) Effective values of $\alpha$ for
with~($\Box$) and without~($\circ$) diffusion.}
\label{fig:w_d}
\end{figure}

Starting from a flat surface at $t=0$, the surface widths
are measured for $L=2^5,\ldots,2^{9}$. The results are shown
in Fig.~\ref{fig:w_d}(a).
They are qualitatively the same as in the absence of diffusion.
The exponents $\alpha$ and  $\beta$ are determined  in the same way as in
Eqs.~(\ref{alpha_eff}) and (\ref{beta_eff}), see
Fig.~\ref{fig:w_d} (b) :
\begin{equation}\label{alpha_beta_d}
\alpha =  0.31(3) , \quad \beta = 0.115 (5) \ .
\end{equation}
The finite size corrections to scaling are again very large.
The exponents are slightly larger than in Eq.~(\ref{alpha_beta}),
but, within the current numerical accuracy we cannot distinguish one from
the other.

We conclude that dissociative dimer equilibrium dynamics
represents a new universality class for interface roughness.
Surface diffusion within terraces is irrelevant and
this new universality class is characterized by the
topological constraint caused by the mod 2 conservation of the number 
of particles
at every height level.

\section{Even-visiting random walks}\label{sec:evrw}

\subsection{the model}

The above numerical study of dissociative dimer type dynamics
clearly indicates that the equilibrium scaling properties
of the interface belong to a different universality class than
conventional monomer type dynamics.
We also identified the most likely  origin of this:
the constraint that the number of particles at each
height level must be preserved modulo 2 in a global non-local manner.
The exact value of the exponent $\alpha$ is difficult to pin
point from the MC results, due to strong corrections to scaling.
To resolve this, we investigate in this section the properties of
a random walk with the constraint that it needs to visit every site
an even number of times before it terminates.
This is the so-called even-visiting random walk~(EVRW).

Consider a random walker on a 1D lattice, which is required to jump
during each time  step one site to the left or the right with equal
probability, $n_{t'+1} = n_{t'} \pm 1$.
$n_{t'}$ denotes the position of the walker at time $t'$.
The walker is demanded to visit every site $n$ an even number of times
after $t$ time steps.

We focus our presentation on the EVRW in one dimension.
The generalization to $d>1$ is straightforward and mentioned when appropriate.
Moreover, it is natural to expand the EVRW into a Q-visiting random
walk~(QVRW)
with the constraint that each site must be visited a multiple of Q-times.
We obtained numerical results for $Q \ge 3$, but since we did not detect any
differences from the  scaling behaviors at $Q=2$~\cite{JD_unpub},
we limit this presentation to EVRW.

\begin{figure}
\centerline{\epsfxsize=84 mm  \epsfbox{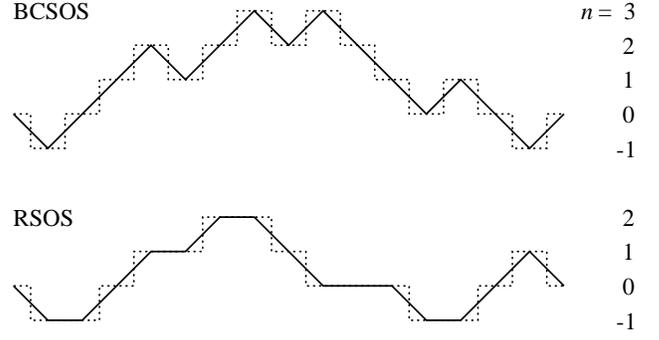}}
\vskip 10 true pt
\caption{EVRW~(solid line) and corresponding surface~(dotted line)
configurations. }
\label{fig:conf}
\end{figure}

The connection with dimer surface dynamics is self-evident.
The probability distribution of EVRW represents the equilibrium Gibbs 
distribution,
i.e., the equilibrium  state of a surface where all
configurations that satisfy the  mod 2 constraint are equally likely.
There is one minor difference between our RSOS dimer model and the above
EVRW.
In the latter the particle is
required to make a hop during every time step, $\Delta n=\pm 1$, while in
the
RSOS dimer dynamics  it is allowed to stay at the same site, $\Delta 
n=0,\pm 1$.
Figure~\ref{fig:conf} shows examples of both.
This so-called body-centered solid-on-solid version of the EVRW is
more compact and converges numerically faster.

\subsection{exact enumerations}

The number of possible space-time configurations of a normal 1D
random walker is equal to $Z_{RW}(t) =2^t$.
The even-visiting constraint excludes most of those walks.
It is of interest to know whether the total number of EVRW's still
scales exponentially as $Z(t) \sim \mu^t$, and if so, whether $\mu$ remains
equal to $2$.
For that purpose, we enumerate all EVRW's that start
and return to the origin ($n=0$) after $t$ time steps, using the
exact (but not closed form) expressions, Eqs.(\ref{S_cicuta}) and 
(\ref{Z0_cicuta})
below, which were developed in Ref.~\cite{Cicuta} in the following manner.

Denote the number of steps to the right~(left) from site $n$ to
$n+1$~($n-1$) by $r_n$~($l_n$).
The number of visits of site $n$ is equal to $v_n = r_n + l_n$
and the sum of all visits is equal to total number of time steps $t = 
\sum_n v_n$.
The return-to-origin condition implies that  $r_n = l_{n+1}$, i.e., that
$v_n = r_n+r_{n-1}= l_{n+1}+l_n$, and that  $l_n$ and $r_n$
must be even for all $n$, such that $t=\sum_n v_n = 2 \sum_n r_n$
is a multiple of $4$ instead of $2$.

Every  walk can be specified by the left boundary
of the walk $n_{\min}$,  and $m$ positive integer variables
$[s_1,\ldots,s_m]$.
The excursion $m$ is defined by the distance between
the right and left boundaries of the walk.
The number of steps from site $n$ to $n+1$ is equal to
$r_n = 2 s_{n-n_{\min}+1}$ with the understanding that $s_{n'}=0$ for
$n'\leq 0$ and $n'>m$.
The number of walks with the same set of positive integers
$[s_1,\ldots,s_m]$ can be readily evaluated and is equal to  ~\cite{Cicuta}
\begin{equation}\label{S_cicuta}
S_{[s_1,\ldots,s_m]} = \frac{t}{2s_1} \prod_{i=1}^{m-1}
\frac{(2s_i + 2s_{i+1}-1)!}{ (2s_{i+1})! (2s_i -1)!}
\end{equation}
for  $m\geq 2$ and is equal to  $S_{[s_1]}=2$ for $m=1$.
The total number of the EVRW's is given by the sum
\begin{equation}\label{Z0_cicuta}
Z^0 (t) = \sum_{m=1}^{t/4} \sum_{\{s_i\}}' S_{[s_1,\ldots,s_m]} \ ,
\end{equation}
where the prime in the second summation denotes the constraint that
$t = 4\sum_i s_i$, and the superscript in $Z^0$  represents the 
return-to-origin condition.

Although analytically exact, this formula still involves infinite sums.
Therefore we must resort to numerical enumerations
to determine the scaling properties. This has to be a finite size 
scaling type analysis
because of the numerical  upper limit for $t$.

\begin{figure}
\centerline{\epsfxsize=84 mm  \epsfbox{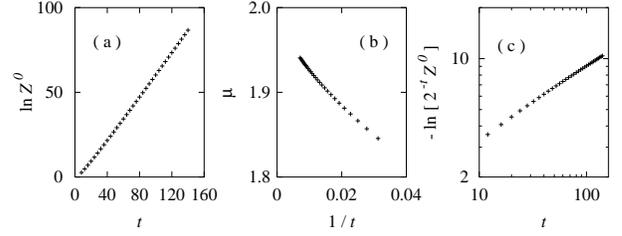}}
\vskip 10 true pt
\caption{Exact enumeration of the total number of EVRW's.
(a) $\ln Z^0$ vs.~$t$, (b) $\mu(t)$
vs.~$1/t$, and (c) log-log plot of $-\ln[2^{-t} Z^0 ]$ vs.~$t$.
}\label{enum_mu}
\end{figure}

In Fig.~\ref{enum_mu}(a), we plot $Z^0 (t)$ as function of time for 
$t\leq 140$.
The linear dependence  in this semi-log plot indicates an exponential form
$Z^0(t) \sim \mu^t$.
Next, we define an effective finite size exponent as
\begin{equation}
\mu(t) = [ Z^0(t) / Z^0(t-4) ]^{1/4}  .
\end{equation}
The corrections to scaling in  Fig.~\ref{enum_mu}(b) are strong,
but a Neville type extrapolation analysis~\cite{Guttmann}
yields
\begin{equation}\label{mu}
\mu = 2.000(2) \ .
\end{equation}
Despite the severe global constraint, the total number of EVRW's scales
asymptotically in the same way as that of normal random walks
with $\mu=2$.

Figs.~\ref{enum_mu}(a) and (b) indicate the presence of
strong corrections to scaling.
They are of an exponential form
\begin{equation}\label{Z_exp}
Z^0(t) \sim 2^t e^{-a t^\theta} \ .
\end{equation}
as shown in  Fig.~\ref{enum_mu}(c). The slope yields
\begin{equation}\label{theta}
\theta=0.34(2).
\end{equation}
In Sec.~\ref{sec:walls}, we  will argue that the 
exponent  $\theta$
is a universal quantity, and equal to the inverse of the dynamic exponent
$z$ of the EVRW; $\theta=1/z$.

\begin{figure}
\centerline{\epsfxsize=84 mm  \epsfbox{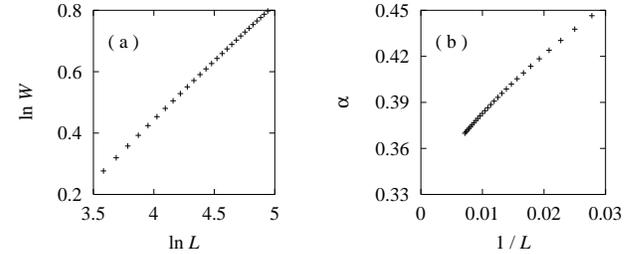}}
\vskip 10 true pt
\caption{Exact enumeration of the EVRW. (a) Surface width $W$.
(b) Effective values of the roughness exponent $\alpha$.}
\label{enum_alpha}
\end{figure}

We also performed an exact enumeration of the finite size scaling of
the width of the EVRW.
All configurations counted in Eq.~(\ref{S_cicuta})
have the same number of visits $v_n$ up to a constant shift  in $n$. Hence,
they all have the same width, $W[\{s_i\}] = ( \bar{n^2}-\bar{n}^2 
)^{1/2}$, with
\begin{eqnarray*}
\bar{n} &=& \frac{2}{L} \sum_{i=1}^m (2i+1) s_i \\
\bar{n^2} &=& \frac{2}{L} \sum_{i=1}^m ( i^2+(i+1)^2) s_i \ .
\end{eqnarray*}
The ensemble averaged surface width
\begin{equation}
W^2 = \frac{1}{Z^0(t=L)} \sum_{m=1}^{t/4} \sum_{\{s_i\}}' W[\{s_i\}]^2
S_{[s_1,\ldots,s_m]} 
\end{equation}
is evaluated  numerically and plotted in
Fig.~\ref{enum_alpha}(a). The roughness exponent $\alpha$, $W\sim L^\alpha$,
is estimated from an effective exponent
\begin{equation}
\alpha(L) = \frac{L}{4}\left(\frac{W(L)}{W(L-4)}-1\right) \ ,
\end{equation}
see Fig.~\ref{enum_alpha}(b).
Again, the convergence is slow, but the Neville
type extrapolation yields
\begin{equation}\label{wijd}
\alpha = 0.327(9) \ .
\end{equation}
Within the numerical accuracy, this result is consistent with those in the
two
dimer growth models (with/without monomer diffusion) of the previous 
section, see
Eqs.~(\ref{alpha_beta}) and (\ref{alpha_beta_d}).

The surface roughness exponent $\alpha$ is simply related to the dynamic
exponent $z$ of the EVRW as
$z=1/\alpha$. So the above numerical result implies that
\begin{equation}\label{z-value}
z = 3.06(8) \ .
\end{equation}
All the results of this section  are checked numerically for $Q=3,4,5$
in the QVRW model.
We find no Q-dependence of the values of scaling exponents~\cite{JD_unpub}.

\subsection{Gaussian distributions}

We performed Monte Carlo simulations  to determine the probability
distribution $P(n,t)$ for the EVRW, i.e., the probability
to start at site $n=0$ and end after time $t$ at site $n$.
This was done by brute force.
We simply generated an ensemble of normal random walks
and trashed the ones that did not satisfy the EVRW condition.
The ratio decreases rapidly.
For example, out of a total of $2\times  10^9$ normal random walks
only about $ 600$ walks satisfy the constraint  at $t=500$.

\begin{figure}
\centerline{\epsfxsize=84 mm  \epsfbox{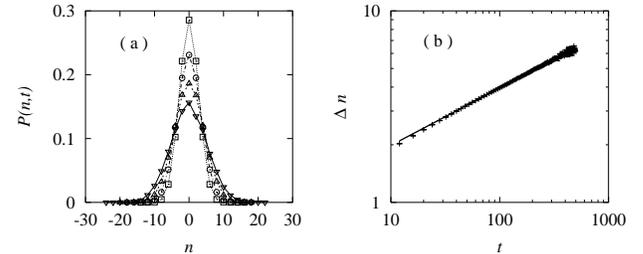}}
\vskip 10 true pt
\caption{Monte Carlo results for the 1D EVRW.
(a) Probability distribution $P(n,t)$ at
$t=32(\Box)$, $64(\circ)$, $128(\bigtriangleup)$, and
$256(\bigtriangledown)$.
(b) Scaling of the spreading  $\Delta n$.}
\label{mc:pdf}
\end{figure}

\begin{figure}
\centerline{\epsfxsize=86 mm  \epsfbox{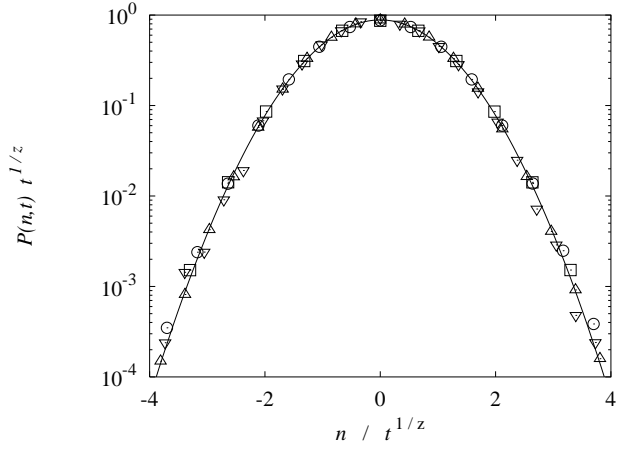}}
\vskip 10 true pt
\caption{Scaling of $P(n,t)$ for the 1D EVRW at 
$t=32(\Box)$, $64(\circ)$, $128(\bigtriangleup)$, and
$256(\bigtriangledown)$ according to Eq.~(\ref{P:scale}).
The best collapse is obtained with $1/z=0.32$.
The scaling function is assumed to be of the form ${\cal
F}(u) = Ae^{-B|u|^\Delta}$. A least square fitting yields $A=0.88$,
$B=0.62$, and $\Delta=1.98$.}
\label{mc:pdf_scale}
\end{figure}

The distribution function is shown in Fig.~\ref{mc:pdf}(a)
and can be assumed to obey the scaling form
\begin{equation}\label{P:scale}
P(n,t) = \frac{1}{t^{1/z}} {\cal F}(n/t^{1/z})
\end{equation}
with $z$ the dynamic exponent.
The best  data collapse  is   obtained   for  $1/z=0.32$,
as shown in Fig.~\ref{mc:pdf_scale}.
This value of $z$ is consistent with
the exact enumeration results of the previous subsection.
It is also consistent with a direct evaluation of the
second moment of the distribution function data, which yields that
\begin{equation}\label{nt0}
\Delta   n =    \left[ \sum_n   n^2 P(n,t)  -  \left(\sum_n  nP(n,t)
\right)^2
\right]^{1/2} \
\end{equation}
scales as $\Delta n \sim t^{1/z}$, with
$z\simeq 3.3$ as shown in Fig.~\ref{mc:pdf}(b).

The functional  form of the scaling function  ${\cal F}$ is a surprise.
It is of the form of ${\cal  F}(u) = A  e^{-B |u|^\Delta}$ ,
as shown in Fig.~\ref{mc:pdf_scale},
with $\Delta\simeq 1.98$. This means that
the probability distribution is Gaussian in nature,
\begin{equation}
P(n,t) = A t^{-1/z} \exp\left[ - B \left(n / t^{1/z} \right)^2 \right] \ .
\end{equation}
This is surprising, because in other models with anomalous surface
roughness, such as Levi flights, the probability distribution is
certainly not Gaussian~\cite{Bouchaud}.

Gaussian distributions with $z=2$ are characteristic for uncorrelated 
random processes.
The appearance of a Gaussian shaped scaling function in the EVRW problem
suggests us to search for an effective  representation of the EVRW in 
which the correlation
effects  somehow transform away, with the possibility for an exact
derivation
of the EVRW dynamic exponent, possibly $z= 3$.
This is the topic of the next section.

\section{Random Walks Coupled to Ising Spins}\label{sec:Ising}

\subsection{defect spreading}

The even-visiting constraint is non-local in time.
To keep track of this constraint in a local way,
we can add an Ising field to a normal conventional random walk,
i.e., a marker  $S_n=\pm 1$ to each site, that keeps track of the
visits in the past.
Initially at time $t'=0$, all spins are prepared in the spin-up state.
$S_n$ flips each time the random walker visits site $n$.
The requirement that all spins are pointing up at $t'=t$,
represents the EVRW constraint.
The generalized distribution ${\cal P}(\{S\};n)_{t}$
contains all the information we need.
$n$ is the location of the random walker at time $t$ and $\{ S \}$
the spin configuration.
${\cal P}(\{+\};n)_{t}$ is the EVRW distribution.

Each down spin at intermediate time $0<t^\prime<t$ represents a defect,
which needs to be healed at a later time. The defect area spreads
in exactly the same way as the width of the conventional random walk,
$W_d\sim t^{\frac{1}{2}}$. We confirmed numerically that the
defect distribution inside this cone is uniform in 1D and 2D.
This allows us to build the following healing time
argument for the value of the EVRW dynamic exponent.

\subsection{defect healing time argument}

Divide the time interval $t$ into two segments,
$\tau_{\text{free}}$ and $\tau_{\text{heal}}$.
For $t'<\tau_{\text{free}}$ the random walker does not feel the constraint,
diffuses freely, and leaves defects behind that
are uniformly spread over a region of size
$\xi\sim \tau_{\text{free}}^{1/2}$.
In order to satisfy the defect-free constraint at time $t$,
the walker stops spreading and starts to
heal defects during the second part of the walk, $t'>\tau_{\text{free}}$.
The typical distance it needs to travel to heal a specific defect is
of order $\xi$, and the time it takes the random walker to do that is of
order
$\tau_1\sim\xi^2$. The total  number of defects $N_d$ is of order $\xi^d$
($d$ is the spatial dimension).
Therefore, the  healing time scales as $\tau_{\text{heal}} \sim N_d 
\tau_1\sim  \xi^{d+2}$.
Putting this all together yields a relation between the final time
and the width $\xi$ of the EVRW.
\begin{equation}
t = \tau_{\text{free}}+\tau_{\text{heal}} \sim \xi^{2}+\xi^{d+2} \ .
\end{equation}
$\tau_{\text{heal}}$ diverges faster than $\tau_{\text{free}}$, so we 
conclude that
\begin{equation}\label{z_healing}
\xi\sim t^{1/(d+2)} \quad \text{and} \quad z = d+2 ~.
\end{equation}
The argument is more subtle in $d>2$ due to the fact that the number
of defects after time $\tau_{\text{free}}$ can not be larger than the 
total number
of time steps, while the volume of the spreading cone, 
$\xi^d\sim \tau_{\text{free}}^{d/2}$,
diverges faster than that.
This implies that in $d>2$ the density of defects
inside the spreading cone does not reach a constant. The number of 
defects inside
$\xi^d$ is only proportional to $N_d\sim \xi^2$ instead of $\xi^d$.
The time to heal one defect, $\tau_1$, however, changes as well.
$\tau_1$ is proportional to the time it takes to travel across the 
spreading cone $\xi$, times the
probability to hit a defect while doing so, which is proportional to 
$\xi^d/N_d$.
The end result is that the healing time still scales  the same as in $d<2$,
\begin{equation}
\tau_{\text{heal}}\sim N_d \tau_1 \sim N_d  ~(\xi^2 \frac{\xi^d}{N_d })\sim
\xi^{d+2} \ .
\end{equation}
We conclude that
$z =d+2 $ in all dimensions.
The value $z=3$ in 1D, is consistent with the
numerical studies of the previous sections.
This derivation is far from rigorous, but has the merit of being
simpler than the ones in the following sections.

The separation of $t$ into two distinct time scales
$\tau_{\text{free}}$ and $\tau_{\text{heal}}$ is artificial.
Consider the average over all possible starting positions of 
the random walker and all possible spin configurations 
with periodic boundary conditions in the time direction (full trace). 
Then, the system becomes translationally invariant in the time direction 
and the two distinct time domains should disappear. However, 
$\tau_{\text{free}}$ is still the natural crossover time scale in the problem.
Consider the EVRW over time interval $t$.
Measure the width of the walk in a smaller time window $t^\prime$ inside
$t$. For very small windows, $t^\prime \ll t$, the even-visiting constraint is invisible, and
the width scales in the same manner as for a normal conventional random walk.
This implies the following crossover scaling form for the width, 
$\xi$, of the EVRW
\begin{equation}\label{scalingform2}
\xi(t^\prime,t)=b \xi(b^{-z_{rw}} t^\prime, b^{-z} t) = t^{1/z} 
G(t^\prime/\tau_{\text{free}}) \ ,
\end{equation}
with $b$ an arbitrary scale factor and $z_{rw}=2$.
$G$ is  the crossover scaling function
and $\tau_{\text{free}}=t^{z_{rw}/z} $ the crossover time scale.

This crossover is important from a surface science perspective.
The time scale $t$ corresponds to the characteristic length scale between
impurities or other surface defects that act as effective lattice cutoffs.
Depending on the experimental set up,
like an X-ray beam width or STM scanning window that 
might be larger or smaller than this, one may measure 
the true asymptotic surface width scaling with 
$\alpha=1/(d+2)$, or the unconstrained value $\alpha=1/2$.

\begin{figure}
\centerline{\epsfxsize=86 mm  \epsfbox{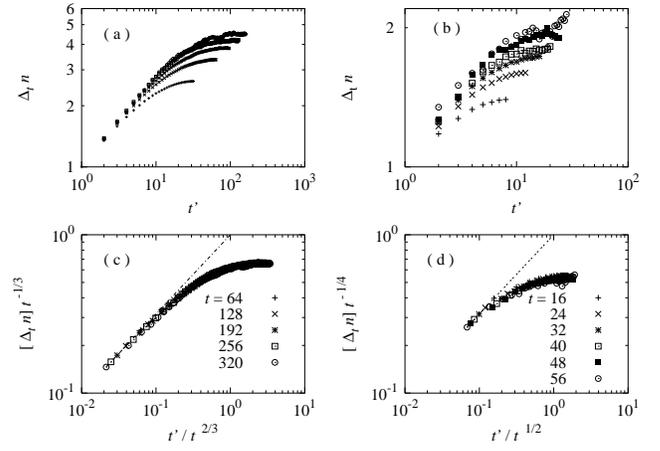}}
\vskip 10 true pt
\caption{Crossover scaling in the EVRW.
(a) and (b) Spreading of the EVRW in $d=1$ and $d=2$.
(c) and (d) Scaling of the data according to Eq.~(\ref{scalingform}).
The broken lines have slope 1/2.}
\label{DelX}
\end{figure}

To illustrate the existence of this crossover time scale,
we measure the spreading of the EVRW's
\begin{equation}\label{nt'}
\Delta_t n_{t^\prime} \equiv \left[ \langle\!\langle n_{t'}^2
\rangle\!\rangle_t -  \langle\!\langle n_{t'} \rangle\!\rangle_t ^2
\right]^{1/2} \ .
\end{equation}
$\langle\!\langle \cdots \rangle\!\rangle_t$ denotes the ensemble
average over the walks that satisfy the even-visiting constraint at time
$t$.
Note that $\Delta n$ in Eq.~(\ref{nt0}) is equal to $\Delta_t n_t$.
The spreading must obey the same type of crossover scaling form as in
Eq.~(\ref{scalingform2}),
\begin{equation}\label{scalingform}
\Delta_t n_{t'} = t^{1/(d+2)} {\cal G}(t'/ t^{2/(d+2)}) \ ,
\end{equation}
Monte Carlo simulations confirm this. We generate EVRW's
over a given time interval $t$ subject to  the return-to-origin constraint,
and record the time trajectories for $0\le t'\le t/2$.
Fig.~\ref{DelX} shows the spreading in (a) $d=1$ and (b) $d=2$.
The crossover behavior is clearly visible in
Figs.~\ref{DelX} (c) and (d).
The data for different $t$ collapse very well.
As expected from Eq.~(\ref{scalingform}),
the scaling function ${\cal G}(u)$ increases as $u^{1/2}$
in the short time region and saturates to a constant in the opposite limit.

\subsection{stochastic spin flip dynamics}

Consider a generalization of the EVRW in which the random walker
flips the spin only probabilistically during each visit.
The spin flips  with probability $e$ or is left
unchanged with probability $f=1-e$.

As in the deterministic EVRW problem, we
require that at time $t$ all spins return to the spin-up position.
A more elegant and equivalent formulation of this is to
require time-like periodic boundary conditions,
because it suffices to demand that all spins at time $t$
return to the same state as at time zero, irrespective of what
that state might be, and the trace over all such initial conditions
leads to periodic time-like boundary conditions.

We call this model the stochastic even-visiting random walk~(SEVRW).
The deterministic EVRW corresponds to $(e,f)=(1,0)$
and the conventional  RW to $(e,f)=(0,1)$.

The purpose of this generalization is two-fold.
On the one hand, it allow us to address the robustness of anomalous 
EVRW diffusion.
On the other hand, and more importantly, there is an exactly solvable
``decoupling point", $e=f=1/2$, where we can evaluate the
anomalous diffusion scaling rigorously, see Sec.~\ref{sec:walls}.

\subsection{non-Hermitian quenched randomness}\label{sec:SEVRW}

The master equation for the probability distribution reads
\begin{eqnarray}
{\cal P}(\{S\};n)_{t+1} &=& \frac{e}{2} \left[ {\cal P}(\{S'\};n+1)_{t} +
{\cal P}(\{S'\};n-1)_{t} \right] \nonumber \\
&+& \frac{f}{2} \left[ {\cal P}(\{S\};n+1)_t + {\cal P}(\{S\};n-1)_t 
\right] \ ,
\end{eqnarray}
where configurations $\{S^\prime\}$ and $\{S\}$ are related as
$S'_{n} = -S_{n}$ and $S'_{m}=S_m$ for $m\neq n$.
This can be cast in state vector  notation,
$|{\cal P}\rangle_{t}  =   \sum_{\{S\},n}  {\cal P}(\{S\};n)_{t} | 
\{S\};n\rangle$,
as
$| {\cal P}\rangle_{{t}+1} = \hat{T} ~ | {\cal P}\rangle_{t}$
with the time evolution operator
\begin{equation}\label{gtt}
\hat{T}  =   \frac{1}{2}  \sum_n   ( e \hat{\sigma}_n^x  +f )
\left[\hat{a}_{n}^\dagger  \hat{a}_{n+1}   +
\hat{a}_{n}^\dagger \hat{a}_{n-1} \right] \ .
\end{equation}
The $x$-components of the Pauli spin operators, $\hat{\sigma}^x$ represent
the spin flips, and the fermion annihilation/creation operators
$\hat{a}$ and $\hat{a}^\dagger$ represent the random walker.
We have only one fermion in the energy band.

The spin part of $\hat{T}$ is easily diagonalized since the
$\hat{\sigma}^x$ do not couple to each other directly.
Perform a  rotation in spinor space
to the  eigenvectors,
$\frac{1}{\sqrt{2}} (|+\rangle \pm |-\rangle)$, of $\hat{\sigma}_n^x$,
and denote the eigenvalues as  $c_n = \pm 1$.
In the rotated spinor basis, the operators $\hat{\sigma}_n^x$ become
$c$-numbers, $c_n$, and the time evolution operator reads
\begin{equation}\label{gT(c)}
\hat{T}(\{c\})   =  \frac{1}{2}\sum_n   ( e c_n   +f)
\left[\hat{a}_{n}^\dagger  \hat{a}_{n+1}   +
\hat{a}_{n}^\dagger \hat{a}_{n-1} \right] \ .
\end{equation}
The initial all-spin up configuration  becomes in the rotated spinor basis
the linear superposition over all possible $\{c_n\}$.
Each $c_n$ is either $+1$ or $-1$ at random and does not evolve in time.
The fermion (random walker) hops on a 1D lattice with randomly
placed defects, the $c_n=-1$ sites.
The spin degrees of freedom transform into quenched random noise in the
hopping
probabilities. The time-periodic boundary conditions for the original spin
variables translate into a quenched average over all defect configurations
distributed uniformly.
The wave function (probability distribution) is
multiplied by a factor $(-e+f)$ each time the fermion visits a defect.
Notice that the probability distribution can be negative when $e>f$
for certain defect configurations.

The generalization to Q-visiting random walks is straightforward.
The eigenvalues become complex, $c_n=\exp{(i2\pi j/Q)}$ with 
$j=1,\dots, Q$,
and $(Q-1)$ different kinds of defects appear with different random
hopping probabilities. This type of generalization
does not lead to any new scaling behavior of the
probability distribution of the random walker
in the asymptotic limit~\cite{JD_unpub}.

The time evolution operator in Eq.~(\ref{gT(c)}) resembles the
Hamiltonian for an electron in a random medium.
One  fundamental difference is that $\hat T$ is non-Hermitian.
The hopping probability from $n$ to $n+1$ is not Hermitian conjugate
to that from $n+1$ to $n$.
Non-Hermitian  random Hamiltonians  arise in  various
areas  of physics.  Stochastic processes, like random walks
in disordered environments,  have non-Hermitian time evolution operators.
Equilibrium systems with quenched
disorder, like vortex line pinning in dirty superconductors~\cite{Hatano_Nelson}
are described in the transfer matrix formulation by a
non-Hermitian  random Hamiltonian.
Delocalization transitions
for such non-Hermitian types of disorder are different in nature
from those in Hermitian systems, see e.g., Ref.~\cite{Feinberg_Zee}.

This relation between  non-Hermitian random Hamiltonians and
the EVRW is not new. It is presented typically starting from the
non-Hermitian
perspective. Our derivation presented above  using the reverse route
has (in our opinion) the advantage of being more transparent.
To be precise,  Cicuta {\it   et al}~\cite{Cicuta}
recently considered  a ``roots  of unity'' model with  Hamiltonian
\begin{equation}\label{cicuta}
H = \sum_n (\hat{b}^\dagger_{n} \hat{b}_{n+1} +
\chi_n \hat{b}^\dagger_{n} \hat{b}_{n-1} )
\end{equation}
where $\hat{b}$, $\hat{b}^\dagger$ is a fermion operator and
$\chi_n=\pm 1$ is the  random variable with a uniform distribution.
They relate this  non-Hermitian random
Hamiltonian to the  deterministic EVRW.
The disordered  average of the trace of $H^t$  generates   the   EVRW
configurations~\cite{Cicuta}.
Alternatively, the similarity transformation $\hat{a}_n =
\zeta_n \hat{b}_n$ with
$|\zeta_n|^2 =1$ maps  the time evolution operator in Eq.~(\ref{gT(c)})
onto Eq.~(\ref{cicuta}) with $c_n  = \zeta_{n}/\zeta_{n+1}$
and $\chi_n = c_n c_{n-1}$.

\subsection{polymers in random media}

In the spin diagonalized form of Eq.~(\ref{gT(c)}), the single fermion
is equivalent to a walker (fermion) in a quenched random environment.
With probability 1/2 each site  $(n)$ is occupied by a defect, $c_n=-1$,
or not, $c_n=+1$. The probability distribution satisfies the recursion
relation
\begin{eqnarray}
{\cal P}(\{c\},n)_{t+1} &=& \frac{1}{2} (ec_n+f) \nonumber \\
&&\left[ {\cal P}(\{c\},n+1)_{t} + {\cal P}(\{c\},n-1)_{t} \right] \
\end{eqnarray}
During each time step, ${\cal P}$ is multiplied with
a factor $\frac{1}{2}$ and with an additional
factor $f-e$ each time the walker lands on a defect site (recall that
$f+e=1$).
This equation  of motion does not preserve probability,
and therefore we can not interpret it as a Master equation.
The random walk nature of the problem is only restored after taking the
quenched average over the $c_n$ randomness.

Instead, we can interpret this equation of motion as
the transfer matrix of a polymer wandering (but not back bending)
on a 2D $(n,t)$-lattice with defect lines
(at specific $n$ along the $t$ direction).
The partition function is equal to
\begin{equation}\label{polymer1}
P(n,t) = 2^{-N_s} \sum_{\{c\}} \sum_{\text{walks}} 2^{-t} e^{-\mu 
\sum_{n^\prime} v_{n^\prime}}
\end{equation}
with $N_s$ the number of lattice sites, $v_n$ the number of times
the polymer visits site $n$ in the specific walk under consideration,
and $\mu=-\log(f-e)$ the energy
associated with hitting a defect line.
The prime in $n^\prime$ represents that we only sum
inside the exponential over defect sites.

The SEVRW interpolates between the normal random walk and the EVRW.
At the random walk point, ($e=0,f=1$),
the defects decouple from the polymer.
At the EVRW point, $(e=1,f=0)$, the summand in the partition function
changes sign each time the polymer hits a defect line.

Next, we can integrate out the defects altogether, because
the order of the two summations, the one over all polymer walks
and the one over all
possible defect line configurations, $\{c\}$, can be interchanged.
(From the polymer perspective, the disorder is annealed, not quenched.)
The trace over all defect configurations leaves us with
\begin{equation}\label{polymer}
P(n,t) =
\sum_{\text{walks}} 2^{-t} \prod_m \left[\frac{1}{2} \left( 1+e^{-\mu 
v_{m}}\right)\right]
\end{equation}
with the product now running over all lattice sites $m\in N_s$.
This leads us back into familiar territory.
The SEVRW problem is now reformulated
as a trace over normal unconstrained RW's, but with Gibbs type weights
giving each walk a different probability  depending on
the number of visits $v_n$ of every site.
We could have started this way, because
at $e^{-\mu}=-1$  Eq.~(\ref{polymer})
counts naturally  only the EVRW,
and at $e^{-\mu}=1$ it counts all RW.
For other values of $\mu$ the walks are weighted in
a more complicated way, except at $e=f$, as we will discuss next.

\section{The Exactly Solvable Point}\label{sec:walls}

\subsection{reflective walls}

At point $e=f$ the SEVRW is exactly solvable. Here
the properties of the walk simplify in a different manner than at
the EVRW point, $e=1$, and the normal RW point, $f=1$.
The generating function representation of Eq.(\ref{polymer})
reduces to
\begin{equation}
P(n,t) =
\sum_{\text{walks}} 2^{-t-N_v}
\end{equation}
with $N_v$ the number of distinct sites visited by that particular random walk.
The total number of walks is equal to
\begin{equation}\label{Zdecoupl}
Z(t)= 2^t\sum_n P(n,t) = \sum_{\text{walks}} e^{-h N_v}
\end{equation}
with $h= \log 2$, and the summation running now over all walks
irrespective  of their end point. $N_v$ is also equal to the distance
between the two extremal points reached by the RW.
It is as if an energy is being assigned to each
RW proportional to its space-time width.

In the formulation of Eq.~(\ref{polymer1}) the polymer is not
allowed to cross defect lines ($\mu$ diverges),
i.e., the problem  factorizes in random sets of polymers on strips with
finite widths.
Similarly, the fermion time evolution operator reduces to
\begin{equation}\label{T0}
\hat{T} = \frac{1}{4} \sum_n (c_n +1)
\left[\hat{a}_{n}^\dagger  \hat{a}_{n+1}   +
\hat{a}_{n}^\dagger \hat{a}_{n-1} \right] \ .
\end{equation}
The hopping probability to cross defect sites, $c_n=-1$, is zero.
The defects act as {\em hard core} walls. The fermion is trapped
and localized between two neighboring defects.
These reflective walls are randomly distributed with a probability
$1/2$ to find one at every site without any spatial correlations.

The probability to find in the quenched average
the fermion within a blocked line segment of length $\xi$
is proportional to $\xi 2^{-\xi}$;
because the probability to randomly place the fermion on a line
segment of length $\xi$ is proportional to $\xi$, and the probability
that such a line segment
exists in the quenched average is proportional to $2^{-\xi}$.
This allows us to calculate several quantities analytically in 1D.

\subsection{total number of walks}
The total number of SEVRW walks
can be reformulated as
\begin{equation}
Z(t) \sim \sum_\xi  \xi 2^{-\xi}~{\cal Z}(\xi,t) ,
\end{equation}
where ${\cal Z}(\xi,t)$ is the number of possible normal random walks
within a line segment of size $\xi$ with reflective boundary conditions.
A heuristic evaluation of ${\cal Z}(\xi,t)$ runs as follows.

For $t<\xi^2$, ${\cal Z}$ grows as ${\cal Z} \sim 2^t$ just like
normal random walks, but after this typical time scale the random walker
begins to hit the boundary. It can only bounce back instead of
having two possible futures (hopping directions).
So compared to a walk in infinite space  without reflective walls, the total
number of walks is reduced by a definite factor each time the walker
hits the wall. During time $t$, the random walker hits the boundary
$\sim t/\xi^2$  times on average. So one expects
\begin{equation}
{\cal Z}(\xi,t) \sim 2^t \exp[-a t/\xi^2] \ ,
\end{equation}
with  $a$ is a constant of $O(1)$. The
total number of configurations then scales as
\begin{equation}
Z(t) \sim \int d\xi \ 2^t \xi \exp(-a t/\xi^2 - \xi \log 2) \ .
\end{equation}
The integral can be evaluated from the method of steepest descent in 
the limit of
large $t$ :
\begin{equation}\label{Z_analytic}
Z(t) \sim 2^t t^{1/2} \exp(-b t^{\theta})
\end{equation}
with $\theta=1/3$ and $b$ a constant.
The maximum contribution comes from $\xi_m\sim t^{1/3}$ and the power-law
correction term follows in second order.

The total number of walks returning to the origin, $Z^0 (t)$,
can be calculated in a similar way.
The return-to-origin constraint reduces ${\cal Z}(\xi,t)$
by a factor of $\xi$. We obtain
\begin{equation}\label{Z_analytic0}
Z^0 (t) \sim 2^t t^{1/6} \exp(-b t^{\theta}) \ .
\end{equation}

\subsection{spreading exponent}
The spreading, $\Delta n(t)$, of the walker can be evaluated as well.
First consider width $w(\xi,t)$ of a random walker trapped
on a line segment of length $\xi$.
Initially, for $t<\xi^2$, the random walker diffuses
normally with
$w(\xi,t)\sim t^{1/2}$, until it realizes it is trapped.
So $w(\xi,t)$ saturates to $\xi$, and  the spreading scales
as $w(\xi,t) = \xi g(t^{1/2}/\xi)$ with $g(x)\sim x$
for small $x$ and $g(x)$ constant for large $x$.
The total spreading is the average of this :
\begin{equation}
\Delta n(t) = \langle w(\xi,t)\rangle =
\frac{\int d\xi {\cal Z}(\xi,t) \xi 2^{-\xi} w(\xi,t)}
{\int d\xi {\cal Z}(\xi,t) \xi 2^{-\xi}}
\end{equation}
We use the method of steepest descent for large $t$, and again
the maximum contribution comes from $\xi_m \sim t^{1/3}$.
This leads to
\begin{equation}\label{nt}
\Delta n(t) \sim t^{1/3} \ ,
\end{equation}
i.e., $z=3$ (since $\Delta n(t)\sim t^{1/z}$),
or after taking the crossover scaling
into account,
\begin{equation}
\Delta n(t) = t^{1/3} g(t^{1/6}) \ .
\end{equation}
The crossover scaling  dies out very slowly at large $t$,
such that the corrections to scaling are large.

\subsection{exponent identity }

We just established  that the width scales as
$\Delta n \sim t^{1/z}$ with $z=3$, see Eq.~(\ref{nt}), and that the
total number of walks has a correction factor
$\exp{[-b t^{\theta}]}$ with $\theta=1/3$, see Eqs.~(\ref{Z_analytic})
and (\ref{Z_analytic0})).
We will demonstrate  now that $\theta=1/z$.

The total number of constrained walks, $Z(t)$, at the decoupling point
is given by Eq.~(\ref{Zdecoupl}).
The average  width of the random walk is equal to
\begin{equation}
\langle N_v[h] \rangle = -\frac{\partial}{\partial h} \ln Z[h] \ .
\end{equation}
Integrating this equation leads to the formal relation
\begin{equation}\label{Z[s]}
Z[h] = Z[0] ~ \exp\left[-\int_0^h \langle N_v[h'] \rangle dh'\right]
\end{equation}
with $Z[0] = 2^t$.
It is reasonable to presume that $\langle N_v[h']\rangle$ is continuous as
function of $h'$.
Then, according to  the mean value theorem, the integral in the exponent
is proportional to $h \langle N_v[{h'}_0]\rangle$ for 
$0<{h'}_0 \le h$.  By setting
${h'}_0= \log 2$, we obtain
\begin{equation}\label{Z_m}
Z(t) \sim 2^t \exp[ - a \langle N_v \rangle] \ .
\end{equation}
$N_v$ is equal to the excursion width of the walks, and proportional
to $\Delta n$. Therefore
\begin{equation}\label{Z_Delta_n}
Z(t) \sim 2^t \exp[ - a \Delta n(t)] \ .
\end{equation}
Hence we conclude that the exponential factor in the partition function
originates from the spreading of the walks and that $\theta=1/z$.

\subsection{universality}

Our numerical results for the EVRW model of the previous sections agree
with all the above exact results at the reflective wall point;
see Eqs.~(\ref{Z_exp}), (\ref{theta}), (\ref{wijd}), and (\ref{z-value}).
This is actually somewhat surprising.

It is relatively easy to argue that the scaling properties in the 
direct vicinity
of the decoupling point $e=f$ should be robust and universal,
with the decoupling point acting as stable
``fixed point" in the sense of renormalization transformations.
At  the decoupling point  the fermion is deflected by the defects,
while at $e\neq f$ it can tunnel through them.
This tunnelling is an exponentially small effect, see Eq.~(\ref{polymer}).
Passing through two defects is equivalent to passing
through only one at a much smaller value of $f-e$,
which means that under a rescaling of the spatial resolution the
renormalized
$e^{-\mu}$ decreases towards zero.

The normal random walk, at  $f=1$, and the
deterministic EVRW, at  $e=1$ mark the natural  horizons
of the basin of attraction of this $e=f$ fixed point.
At these points, $e^{-\mu}$ becomes equal to $\pm 1$, respectively.
So it remains surprising that the scaling properties
of the deterministic EVRW are the same as in the
reflective wall model.

The following intuitive derivation of
Eq.~(\ref{Z_Delta_n})  sheds some light on this.
We expect that the total number of walks in every type of SEVRW
is proportional to the total number of normal random walks $2^t$,
times the probability that the Ising spin configuration satisfies
the global constraint.  At the $e=f$ point, the Ising spins flip
randomly when their sites are visited. Therefore all spins inside
the spreading cone are randomized completely and lack any spatial
correlations. This means that the  probability to find all Ising spins
pointing up is proportional to $\exp [-a\Delta n]$, which
confirms Eq.~(\ref{Z_Delta_n}).

The extension of this argument to general SEVRW and the EVRW point in
particular, requires that the distribution of down spins is still uniform
and that the  spin-spin correlations are short-ranged in the large $t$ limit.

At the EVRW point, the random walker flips the spin at every visit.
For large $t$, it is very likely  that the number of visits to every site
inside the spreading  region is  even or odd with equal probability;
we checked this numerically.
Spin-spin correlations are the strongest at the EVRW point,
but since this is a 1D chain of Ising spins it is very unlikely that they
can develop  long range order of any type.
We numerically measure the spin-spin correlation function,
$\langle S_n S_{n+r}\rangle$,
and find exponential decay in the spatial direction; the correlation length
saturates to a finite value for large $t$~\cite{JD_unpub}.
This explains why Eq.(\ref{Z_Delta_n}) still holds at the EVRW point.

\section{Lifshitz Tails in Random Hamiltonians}\label{sec:Lifshitz}

\subsection{density of states}

Let's return to the fermion time evolution operator Eq.~(\ref{gT(c)}),
and examine the same scaling issues from that perspective.
The number of walks
$Z^0(t)$ returning to the origin after
$t$ steps ($n=0$) and satisfying the EVRW constraint can be written as
\begin{eqnarray}\label{disorder-dos}
Z^0(t) &=& 2^t 2^{-N_s} \sum_{\{c\}} \langle 0 | (\hat{T}(\{c\}))^t | 0 \rangle 
\nonumber \\
&=& 2^t \int dE \rho(E) E^t \ , \label{Z_rho}
\end{eqnarray}
where $E$ is an eigenvalue of ${\hat T}$ and the disorder-averaged
density of states is denoted by $\rho(E)$. Since the operator is
non-Hermitian, $E$ is a complex number and the integration
runs over the complex $E$ plane. Eigenstates near the band
center are rather well documented for this type of non-Hermitian
random Hamiltonians~~\cite{Hatano_Nelson,Feinberg_Zee}. However,
we need to focus on the eigenstates near the band edge
(at small wave numbers) since there is only one fermion
in the system  and our interests lie with its long time behavior.

The nature of the eigenstates near the band edge,
is rather well-known for Hermitian random systems.
The density of states  $\rho(E)$ of
these edge states exhibits an essential singularity,
known as a Lifshitz tail~\cite{Luttinger}.
We review here an intuitive argument for
the existence of  Lifshitz tails and extend it
to the non-Hermitian random SEVRW model.

\subsection{Lifshitz tails}

Consider a 1D free fermion Hamiltonian with bond disorder:
\begin{equation}
H =-\frac{1}{2} \sum_{n,m} t_{n,m} \hat{a}_n^\dagger \hat{a}_m \ ,
\end{equation}
where $t_{n,m}=t^*_{m,n}$ are random hopping amplitudes between
sites $n$ and $m$. This Hamiltonian is Hermitian.
For simplicity, assume that the $t_{n,m}$
are nonzero only for pairs of nearest neighbor sites
and take only the values $1$ and
$b$ $(0<b<1)$ with equal probability.

Without disorder, with all $t_{n,m} = 1$, the energy band is trivial,
$E = -\cos k$, with  uniformly  distributed  wavenumbers, $\delta k=2\pi/L$,
in the range $(-\pi<k\le \pi)$.
The states near the lower band edge, $k\simeq 0$
describe the large length scale behavior, and
the density of states diverges as a power-law
with the familiar van Hove singularity
\begin{equation}
\rho(E)\sim |\Delta E|^{-1/2} \ ,
\end{equation}
in terms of  $\Delta E = E-E_{edge}$.

The eigenstates  become localized in the presence of disorder.
The probability to find a pure domain, i.e., a connected string of 
$t_{n,m}=+1$,
of size $\xi$ decreases exponentially as $2^{-\xi}$.
The crucial feature behind Lifshitz tails
is that the states extending across the boundaries of pure domains
do not contribute to the density of states near the edge,
even in the presence of small tunneling probabilities ($b>0$).
In that case, the energy levels $E_\ell$ in each segment 
are similar to those of a free particle in a box of size $\xi$, i.e.,
$\Delta E= E(k)-E_{edge}\simeq  k^2/2$, with wave number spacing
$\delta k=2 \pi/\xi$;
or, phrased in terms of the domain size
$\xi$, $|\Delta E_\ell | \sim (\ell/\xi)^2$ for low-lying eigenstates
with $\ell=1,2,\cdots$.

The distribution of first excited states $\rho_1(E)$
between energy $E$ and $E+dE$ is proportional
to the probability to find a pure domain segment with a
size between $\xi$ and $\xi + d\xi$, which is $\rho_1 (E) dE \sim 2^{-\xi} 
d\xi $.  Therefore,
$\rho_1 (E) \sim |\Delta E|^{-3/2} \exp [-a|\Delta E|^{-1/2}]$.
Similarly, for the $\ell$-th level,
$\rho_\ell (E) \sim \ell |\Delta E|^{-3/2} \exp [-a\ell|\Delta E|^{-1/2}]$.
The total density of states is the sum over all levels, but
near the band edges, the contributions from higher
levels yield only corrections to scaling.
Hence the density of edge states is of the form
\begin{equation}\label{rhodens}
\rho (E) \sim |\Delta E|^{-3/2} \exp [-a|\Delta E|^{-1/2}] \ .
\end{equation}
This exponential factor in  the density of states near
the band edge is known as a Lifshitz tail.
Rigorous calculations confirm its existence ~\cite{Luttinger}.
Moreover, the tails exist also in higher dimensions in the form of
\begin{equation}
\rho (E) \sim \exp [-a|\Delta E|^{-d/2}] \ ,
\end{equation}
because, roughly speaking, $\rho_1$ then scales as  $2^{-\xi^d}$.

\subsection{Hermitian SEVRW model}

Let's now generalize this to negative hopping amplitudes. This
may not be useful to real fermions in disordered media, but is
helpful to understand SEVRW's. Consider a Hermitian analogue
of the SEVRW model
\begin{equation}\label{hermSEVRW}
T = \frac{1}{2} \sum_{n,m} (ec_{n,m} +f) \hat{a}_n^\dagger \hat{a}_m \ ,
\end{equation}
where the sum is over nearest neighbor pairs and $e=1-f$
with $0 \le e, f \le 1$. The random variable
$c_{n,m}=c_{m,n}$ can be either $+1$ or $-1$ with equal probability.
So the hopping amplitude $t_{n,m}=ec_{n,m}+f$ is either $+1$ or $(-e+f)$,
and can be negative for $e>f$. The conventional Lifshitz tail argument
applies
to $e<f$.

Similar to our earlier discussions, the $c_{n,m}$ can be regarded
as eigenvalues of Ising-type spin flip operators $\hat{\sigma}_{n,m}^x$.
Unlike before, these Ising spins live on the bonds instead of the sites.
$e$ is the spin-flip probability when the walker (fermion) passes through
the bond. Point $(e,f)=(0,1)$  corresponds to the normal RW model just like
in
the SEVRW model. However, there is an important difference between
the Hermitian and the non-Hermitian versions.  The Hermitian formulation
satisfies a self-duality relation between $(e,f)$ and $(f,e)$.
The following transformation on the creation/annihilation operators
\begin{equation}
\hat{b}_n = \hat{a}_n \prod_{p=1}^{n-1} c_{p,p+1} 
\end{equation}
maps $(e,f)$ onto $(f,e)$.
Therefore, the two limiting points $(e,f)=(0,1)$ and $(1,0)$ must
correspond both to the normal unconstrained RW.
There is an even-visiting condition at  point $(e,f)=(1,0)$,
but it is imposed on the bonds.
Unlike the site version, the bond
constraint is automatically satisfied by all
normal random walks returning to the origin. 
Consider a simple walk as example:
walk ten steps to the left and then all the way back.
When the RW turns around, it leaves a defect behind at the extremal
point, in the site version but not in the bond version.
On its way back it repairs all defects left behind during the
first part of the journey, in both the site and bond versions.
So in the bond version, all defects are automatically repaired.
In the non-Hermitian version (the original SEVRW model) the self-duality
does not exist and $(e,f)=(1,0)$ is the anomalous EVRW problem.

At the decoupling point $e=f=1/2$, the 1D chain of Eq.~(\ref{hermSEVRW})
breaks up completely into randomly distributed finite segments,
just like before in the non-Hermitian SEVRW. The
hopping amplitudes $t_{n,m}$ are either $+1$ or $0$.
These disconnected sections correspond to the pure domains in the
Lifshitz argument and all states are completely localized within
those  sections. The Lifshitz tail argument is exact at the decoupling
point.

The partition function, Eq.~(\ref{disorder-dos}),
is easily evaluated with the method of steepest descent as
\begin{eqnarray}\label{Z0-lif}
Z^0(t) &\sim& 2^t \int dE \  |\Delta E|^{-3/2} \exp[-a |\Delta 
E|^{-1/2} - t|\Delta E|] \nonumber \\
      &\sim& 2^t t^{1/6} \exp(-b t^{1/3}) \ .
\end{eqnarray}
As expected, we have exactly the same formula as
in Eq.~(\ref{Z_analytic0}) for the non-Hermitian decoupling point.
Again, the  dynamic exponent is equal to $z=3$ in 1D.

In higher dimensions, the Lifshitz tails are of the form
\begin{equation}\label{lifshitz_d}
\rho(E) \sim \exp[-a |\Delta E|^{-d/2}] \ .
\end{equation}
and therefore the partition function is  proportional to
\begin{equation}
Z^0(t) \sim (2d)^t \exp[ - a t^{d/(d+2)}] \ .
\end{equation}
Recall from Eq.~(\ref{Z_m}) that
\begin{equation}\label{Z_mm}
Z(t) \sim Z^0(t) \sim (2d)^t \exp[ - a \langle N_v \rangle] \ ,
\end{equation}
where $\langle N_v \rangle$ is the average number of distinct sites visited
by the constrained random walker after $t$ time steps.
Comparing these two equations yields
\begin{equation}
\langle N_v \rangle \sim t^{d/(d+2)} \ ,
\end{equation}
and that every site  is visited $t^{2/(d+2)}$ times on  average.
This implies that $\langle N_v \rangle$ simply scales with the spreading
volume $(\Delta n)^d\sim t^{d/z}$. Therefore,
\begin{equation}\label{Z(d)_con}
Z(t) \sim Z^0(t) \sim (2d)^t \exp [{-a t^{d/z}}] \ ,
\end{equation}
with
\begin{equation}
z = d+2 \ .
\end{equation}
This is the same result as obtained from the healing time argument for the
EVRW, Eq.~(\ref{z_healing}).

\subsection{Lifshitz tails in the EVRW}

The Lifshitz tail argument also applies to the SEVRW
time evolution operator, Eq.~(\ref{gT(c)}).
Since $\hat{T}(\{c\})$ is non-Hermitian, the
density of states is defined in the entire complex $E$ plane.
We focus here on the EVRW point $(e,f)=(1,0)$ where the distribution of
states has a special symmetry property~\cite{Cicuta}.
Apply the similarity transformation
$\hat{a}_{n} = e^{-i\pi/2} \hat{b}_{n}$
to the even sites and leave the odd sites invariant, $\hat{a}_n=\hat{b}_n$.
$\hat{T}(\{c\})$ transforms to $e^{-i\pi /2} \hat{T}(\{c'\})$
with $c'_n = c_n~(-c_n)$ for even~(odd) $n$.
Note that the disorder $\{c'\}$ and  $\{c\}$ have the same 
distribution.
Therefore, one obtains
\begin{equation}\label{symm_rho}
\rho(E) = \rho(e^{i\pi/2} E) \ .
\end{equation}
This symmetry implies that there exist four Lifshitz tails,
along the rays of $\arg (E) = j\pi/2$ with $j=0,1,2,3$ at $|E|=1$.
Each tail contributes equally to the partition function $Z^0 (t)$ apart
from a phase factor $\exp (ij\pi t /2)$ originating from the energy 
eigenvalue at each edge.
So $Z^0(t)$ is equal to Eq.~(\ref{Z0-lif})
multiplied by the constant $\sum_{j=0}^3 \exp (ij\pi t /2)$.
The latter is nonzero only when $t$ is a multiple of 4,
which is trivially true for EVRW's that return to the origin.
We conclude that the dynamic exponent for the non-Hermitian case
is again $z=3$ in 1D and $z=d+2$ in general dimensions.

Finally, we can generalize to  Q-visiting random walks.
The analogue of  Eq.~(\ref{gT(c)})
for $d$ dimensional QVRW's is the time evolution operator
\begin{equation}\label{Td(c)}
\hat{T}(\{c\}) = \frac{1}{2d} \sum_n \sum'_m c_n \hat{a}^\dagger_n
\hat{a}_m
\end{equation}
where $n$ is a site of a $d$ dimensional hypercubic lattice and
the primed sum runs over nearest neighbor sites of given $n$.
The random variable $c_n$ takes equally likely the values $\exp(i2\pi j/Q)$
with $j=1,\dots,Q$.

The density of states has the symmetry property
$\rho(E)=\rho(e^{i\pi/Q}E)$,
following the generalized similarity transformation
$\hat{a}_n=e^{-i\pi/Q} \hat{b}_n$ applied to one
sublattice and leaving the others unchanged, $\hat{a}_n = \hat{b}_n$.
Through this transformation,  $\hat{T}(\{c\})$
picks up a phase factor $e^{-i\pi/Q}$.  There are $2Q$ Lifshitz
tails, along the rays with $\arg (E) = j\pi/Q$ for $j=0,\dots,2Q-1$ at
$|E|=1$.
Each tail contributes equally to the partition function $Z^0 (t)$ except
for the same type of trivial phase factors as in the EVRW ($t$ is now 
a multiple of $2Q$).
The rest of the story is the same as for the EVRW, and the
results are identical.

\section{Summary and Discussion}\label{sec:summ_diss}

In this paper we have investigated the scaling properties of even-visiting
random walks. The number of visits to each site by the random walker is
required to be a multiple of 2. This is a global constraint that leads to
anomalous diffusive motion, of a novel type
compared to more conventional ones such as  Levi flights and 
correlated random walks.
Using exact enumerations and Monte Carlo simulations, we find that
the dynamic exponent is equal to $z=3$ in 1D.
Surprisingly, the probability distribution is not a stretched
Gaussian (as for the other types of  anomalous diffusion) but a simple
Gaussian (with an anomalous value of $z$).
We devise an healing time argument
which suggests that $z=d+2$ in $d$ dimensions.
These results are verified numerically in 1D and 2D.

We embed the even-visiting random walk into an Ising type environment,
with an Ising spin at every site,
where the random walker flips the spin at the site where it lands 
during each visit.
Diagonalizing the spin sector of the master equation, translates
the EVRW into a free fermion problem with quenched randomness.
The time evolution operator takes the form of a non-Hermitian
random-bond free fermion Hamiltonian.
This leads naturally to the formulation of a generalization, stochastic
even-visited random walks (SEVRW).
The master equation for SEVRW  can be reinterpreted as the partition 
function of
a polymer fluctuating in an environment of randomly placed  defect lines,
and, after integrating out the randomness, as the equilibrium partition function
of a  polymer with an
energy proportional to the width of the polymer configurations.

The SEVRW model has a trivially exactly soluble point,
the decoupling point where the polymer can not cross defect lines.
At that point we can show rigorously that $z=3$ in 1D.
Moreover, this point acts as a stable fixed point in a renormalization
transformation type sense in the SEVRW as a whole, such that $z=3$ 
is valid in general.
In the fermion interpretation, the same
asymptotic anomalous diffusive properties of the EVRW
determine the spectral properties of the non-Hermitian Hamiltonian near
the band edge, in terms of so-called  Lifshitz tails.
This confirms that $z=d+2$.

The anomalous roughness we observed numerically
in 1D surfaces described by dissociative dimer-type dynamics
was the starting point and motivation of this study.
Such interfaces provide possible experimental realizations of EVRW's,
like the roughness of steps  on vicinal surfaces where the dynamics
only allow  attachment/detachment in the form of diatomic molecules.

The scaling we found here is very robust. For example,
they also apply to Q-mer type growth models.
We established that the relation $z=d+2$ remains valid
for all  values of Q in the Q-visiting random walk generalization of
EVRW's, and that the probability distribution still takes a Gaussian form.

Random walks are a generic type of stochastic process.
It will be very interesting to search for more novel types of scaling
originating from RW's subject to global constraints.

During the final stages of preparing this manuscript, Bauer, Bernard, and
Luck posted a preprint on the cond-mat archive~\cite{Luck} exploring
the same type of connections between the EVRW and Lifshitz tails.
Their results overlap only partially with the research presented here.

\begin{acknowledgements}
We thank Pil Hun Song and Jysoo Lee for useful discussions.
This work was supported by grant No.~2000-2-11200-002-3 from the Basic
Research Program of KOSEF, and by the
National Science Foundation under grant DMR-9985806.

\end{acknowledgements}

\end{multicols}
\end{document}